\documentclass{article}
\usepackage{graphicx}
\usepackage{hyperref}
\usepackage{amsmath}
\usepackage{cite}

\def\half{{1 \over 2}}
\def\beq{\begin{equation}}
\def\eeq{\end{equation}}
\def\ba{\beq\begin{array}{l}}
\def\ea{\end{array}\eeq}
\def\be{\ba}
\def\ee{\ea}

\def\p{\partial}

\def\g{{ \gamma}}
\def\d{{\delta}}
\def\t{{ \theta}}
\def\e{{\epsilon}}
\def\v{{\varphi}}

\def\cl{{\cal L}}

\def\im{${\cal I}^-$}

\def\sc{$\Sigma_c$}

\def\cO{\mathcal{O}}

\def\x{\hat x}
\def\P{\hat P}
\def\q{\hat q}
\def\r{\hat r}
\def\t{\hat \tau}
\def\v{\hat v}
\def\brownyork{Brown-York}

\def\hm{${\cal H}^-$}
\def\hp{${\cal H}^+$}
\def\hydrolimit{hydrodynamic limit}

\newcommand{\captionfonts}{\footnotesize}

\makeatletter  
\long\def\@makecaption#1#2{%
  \vskip\abovecaptionskip
  \sbox\@tempboxa{{\captionfonts #1: #2}}%
  \ifdim \wd\@tempboxa >\hsize
    {\captionfonts #1: #2\par}
  \else
    \hbox to\hsize{\hfil\box\@tempboxa\hfil}%
  \fi
  \vskip\belowcaptionskip}
\makeatother   

\begin{document}\begin{titlepage}
\centerline{}
\vskip 2in
\begin{center}
{\Large{\bf  FROM NAVIER-STOKES TO EINSTEIN}}
\vskip 0.5in
{Irene Bredberg, Cynthia Keeler, Vyacheslav Lysov
and Andrew Strominger}
\vskip 0.3in
{\it Center for the Fundamental Laws of Nature,
Harvard University\\
Cambridge, MA, 02138

}
\end{center}
\vskip 0.5in
\begin{abstract}

\end{abstract}
We show by explicit construction that for every solution of the incompressible Navier-Stokes equation in $p+1$ dimensions, there is a uniquely  associated  ``dual'' solution of the vacuum Einstein equations in $p+2$ dimensions.  The dual geometry has an intrinsically flat timelike boundary segment $\Sigma_c$ whose extrinsic curvature is given  by the stress tensor of the Navier-Stokes fluid.  We consider a ``near-horizon'' limit in which $\Sigma_c$ becomes highly accelerated.  The near-horizon expansion in gravity is shown to be mathematically equivalent to the hydrodynamic expansion in fluid dynamics, and the Einstein equation reduces to the incompressible Navier-Stokes equation.   For $p=2$, we show that the full dual geometry is algebraically special Petrov type II.  The construction is a mathematically  precise realization of suggestions of a holographic duality relating fluids and horizons which began with the membrane paradigm in the 70's and resurfaced recently in studies of the AdS/CFT correspondence.

\end{titlepage}

\setcounter{page}{1}
\pagenumbering{arabic}

\tableofcontents

\newpage
\section{Introduction}

The Einstein equation
\beq \label{eew}{G}_{\mu\nu}=0,\eeq
and  the incompressible Navier-Stokes equation
\beq\label{nsw}  \dot v_i - \eta \p^2v_i+\p_iP+v^j\p_jv_i=0, \eeq
 have long played central roles in both mathematics and physics. The Einstein equation universally governs the long-distance behavior of essentially any gravitating system, while the incompressible Navier-Stokes equation universally governs the hydrodynamic limit of essentially any fluid.
They both display a rich non-linear structure which has been a continual source of interesting surprises yet remains, centuries after their discovery, incompletely understood.  It is the purpose of this paper to give a mathematically precise
relationship between (\ref{eew}) and (\ref{nsw}) and their solutions, and thereby provide a hopefully useful bridge between the two subjects. For example cosmic censorship could be related to global existence for Navier-Stokes or the scale separation characterizing turbulent flows related to radial separation in a spacetime geometry, see e.g. \cite{min,Oz:2010wz}.

Hints of such a connection, summarized in the next section, have surfaced in various forms over the last three decades \cite{thesDamour,Damsurf,Price,Jacobson:1995ab,Bousso:1999xy,Policastro:2001yc,Policastro:2002se,Kovtun:2003wp,Karch:2003zd,Kovtun:2004de,Gourgoulhon:2005ng,Kovtun:2005ev,Hubeny:2009zz,Bhattacharyya:2008jc,Bhattacharyya:2008ji,Eling:2008af,Bhattacharyya:2008mz,Fouxon:2008tb,Gupta:2008th,Fouxon:2008ik,Eling:2009pb,Eling:2009sj,Paulos:2009yk,Padmanabhan:2009vy,Eling:2010vr,bkls}, see \cite{Son:2007vk,Damour:2008ji,Rangamani:2009xk,Padmanabhan:2009vy,Hubeny:2010wp} for reviews. In particular, excitations of a black hole horizon dissipate very much like those of a fluid \cite{Hawking:1972hy,Hartle:1973zz,Hartle:1974gy,thesDamour,Price,Gourgoulhon:2005ng, Damsurf,bkls,Kovtun:2004de}, and there has been recent discussion of a holographic duality relating black holes and fluids \cite{Hubeny:2009zz,Bhattacharyya:2008jc,Eling:2009pb,min,Kovtun:2003wp,Policastro:2002se,Policastro:2001yc,Kovtun:2005ev,Bhattacharyya:2008mz,Bhattacharyya:2008ji,Gupta:2008th,Eling:2009sj,Kovtun:2004de}. Inspired by these suggestions, in this paper we explicitly construct a map from solutions of the nonlinear incompressible Navier-Stokes equation to solutions of the nonlinear Einstein equation.  A key ingredient is the imposition of boundary conditions which, in a sense to be defined, isolate the horizon dynamics from the rest of the gravitational dynamics and thereby reduce equation (\ref{eew}) to equation (\ref{nsw}).

Our basic construction is roughly as follows. We begin with the region of $p+2$-dimensional Minkowski space inside a hypersurface \sc\ given by an equation of the form $x^2-t^2=4r_c$. \sc\ is intrinsically flat (being the translation of an hyperbola in the t-x plane along the remaining $p$  spatial directions), but has an extrinsic curvature linked to the constant
acceleration $a= 1/ \sqrt {4 r_c}$.  It  asymptotes to its future horizon \hp\ which is the null surface $x=t$. We then study the effect of finite perturbations of the extrinsic curvature of \sc\ while keeping the intrinsic metric flat. These generically lead, when evolved radially inward with the Einstein equation, to singularities on \hp. The special ones which are smooth on \hp\ are analyzed in the hydrodynamic ``$\e$-expansion'', which is a nonrelativistic, long-wavelength expansion and, importantly, keeps terms that are nonlinear in the size of the perturbation. It is found that tensor and scalar modes of the metric decouple in this limit and the remaining degrees of freedom are vector modes governed by the Navier-Stokes equation in $p+1$ dimensions.  We present (equation (\ref{emet}) below) the $p+2$-dimensional solution of the Einstein equation through third order in the hydrodynamic expansion parameter $\e$. The first term is flat space. The second and third terms are algebraically constructed from the velocity field $v^i$ and pressure $P$ of an incompressible fluid. The nonlinear spacetime Einstein equation then reduces to the nonlinear incompressible Navier-Stokes equation for the pair $(v^i, P)$.

This result is already interesting and non-trivial, but the fact that the Navier-Stokes arises when the
geometric variables are subject to the same kind of expansion used in fluid dynamics might have been anticipated. A deeper connection appears when we consider an alternate expansion in which, instead of going to long distances, we take the acceleration of \sc\ to infinity. This is a near-horizon limit since it pushes \sc\ towards its horizon \hp. We then show that, after a constant overall rescaling of the metric,  {\it the near-horizon expansion is mathematically identical to the hydrodynamic expansion}. Hence the solutions of the Einstein equation (constrained by the boundary conditions of a flat metric on \sc\ and smoothness on \hp) in this near-horizon expansion are in one-to-one correspondence with solutions of the incompressible Navier-Stokes equation. This then is the precise mathematical sense in which horizons are incompressible fluids.

It is possible that the ultimate origin of this relation is a deep and exact holographic duality relating (among other things) quantum black holes to fluids as has been suggested by string theoretic investigations. However in this paper we have concentrated on simply establishing the mathematical relationship between (\ref{eew}) and (\ref{nsw}) in a manner which makes no assumptions about or reference to this tantalizing possibility.

This paper is organized as follows. In section 2 we briefly describe precursors of our construction going back to the 70's.  Section 3 briefly reviews the hydrodynamic expansion in the study of fluids, and the emergence of the incompressible Navier-Stokes equation in the hydrodynamic limit. In section 4 we specify the boundary conditions, explained roughly above, used to isolate horizon dynamics. In section 5 we present the general solution of the nonlinear Einstein equation with these boundary conditions through the first three orders in the hydrodynamic expansion, and show that the first nontrivial term corresponds to the velocity field of an incompressible fluid. We also discuss the geometric analog of forcing the fluid, argue for uniqueness, and discuss the possible formation of black hole type singularities. Section 6 presents a simpler form of the metric and shows that, up to an overall rescaling and after an appropriate coordinate transformation,  it depends only on the product of the leading-order acceleration of \sc\ and the hydrodynamic expansion parameter $\e$. In section 7 we show that the geometries are, through the order constructed, of a special type known in four dimensions as Petrov type II. This may enable a connection of the present work with the large literature on algebraically special spacetimes \cite{petr, exac,Milson:2004jx}. Finally in section 8 we demonstrate, using the simplified metric of section 6, the equivalence of the hydrodynamic and near-horizon expansions.

\section{Relation to previous work}

The first suggestion of a relation between horizon and Navier-Stokes dynamics
appears in the prescient thesis of Damour \cite{thesDamour}. This work contains an expression now known as the Damour-Navier-Stokes equation \cite{Gourgoulhon:2005ch} governing the geometric data on any null surface. Although tantalizingly similar, it is not quite the Navier-Stokes equation as it has too many variables (eliminated herein by an appropriate boundary condition) and an extra nonlinear term. To get precisely Navier-Stokes we found it necessary to consider the near-null limit of a highly accelerated timelike surface. Such a surface was introduced by Price and Thorne \cite{Price} and coined the stretched horizon (analogous to our \sc). Although similar in spirit their limit is slightly different from ours. They obtain a compressible fluid with a negative bulk viscosity and an extra term not present in Navier-Stokes.  This approach was developed into the membrane paradigm and is reviewed in the book \cite{Thorne:1986iy}.  Much more recently Policastro, Starinets and Son \cite{Policastro:2002se} made the striking observation, in the context of the AdS/CFT correspondence, that the dissipative behavior of a large black hole in AdS agrees with
that of the hydrodynamics of the holographically dual CFT.  This observation has been developed in many directions \cite{Policastro:2002tn,Kovtun:2003wp,Kovtun:2004de,Saremi:2007dn,Son:2007vk,Baier:2007ix,Brustein:2008cg,Rangamani:2008gi,
Mia:2009wj,Hartnoll:2009sz,McGreevy:2009xe,
Sachdev:2010ch,Hubeny:2010ry,Faulkner:2010jy}. Although far from obvious at first glance, these results from AdS/CFT are compatible with, and in some cases equivalent to,
the earlier results from pure gravity \cite{ Eling:2009pb,bkls,Starinets:2008fb,Iqbal:2008by,Kovtun:2003wp}. In the AdS/CFT context Bhattacharya, Minwalla and Wadia \cite{min,Eling:2009pb} showed that in asymptotically AdS spacetimes at finite temperature, the asymptotic AdS boundary data is governed in a hydrodynamic limit by the Navier-Stokes equation. They use the Navier-Stokes data to construct a bulk solution of the Einstein equation with negative cosmological constant.  Our dual bulk geometry in equation (\ref{emet}) is a refinement of expression (4.4) in \cite{min} in which the cosmological constant is taken to zero and the boundary can be pushed to any radius - in particular to the interesting near-horizon region. Finally we rely heavily on our previous paper \cite{bkls} which solves the linear case within the framework adopted herein.

\section{The \hydrolimit~and the $\e$-expansion}
The incompressible Navier-Stokes equation has a well-known scaling symmetry which is important in the following and briefly reviewed here. Let the pair $(v_i,P)$ solve the incompressible Navier-Stokes equation
\beq\label{nse}
\p^i v_i=0,\qquad
\p_\tau v_i-\eta \p^2v_i+\p_iP+v^j\p_j v_i =0,
\eeq
where $\eta$ is the kinematic viscosity and $i=1,...p$. Now consider a family of pairs $(v^\e_i,P^\e)$ in which frequencies and wavelengths  are non-relativistically dilated
and amplitudes scaled down by the parameter $\e$:
\begin{align}\label{efd}
v^\e_i(x^i,\tau)&=\e v_i(\e x^i,\e^2 \tau),\\
P^\e_i(x^i,\tau)&=\e^2P (\e x^i,\e^2 \tau). \notag
\end{align}
It is easy to check that (\ref{nse}) directly implies
\beq
\p_\tau v^\e_i-\eta \p^2v^\e_i+\p_iP^\e+v^{\e j}\p_j v^\e_i 
=0.
\eeq
Hence (\ref{efd}) generates from the original solution  a family of solutions parameterized by
$\e$.

In real fluids there are always corrections to the Navier-Stokes equation. Galilean invariance requires that these vanish for constant $v_i$.  Typical corrections are for example of the form
\beq
\p_\tau v_i-\eta \p^2v_i+\p_iP+v^j\p_j v_i +v^kv^j\p_k\p_jv_i +\p_\tau^2v_i =0.
\eeq
If $(v_i,P)$ obey this equation, the rescaled quantities obey
\beq
 \p_\tau v^\e_i-\eta \p^2v^\e_i+\p_iP^\e+v^{\e j}\p_j v^\e_i 
+\e^2\bigl(v^{\e k}v^{\e j}\p_k\p_jv^\e_i +\p_\tau^2v^\e_i) =0.
\eeq
 The limit $\e \to 0$ is the \hydrolimit. In this limit these corrections become irrelevant. Similarly the speed of sound goes to infinity and compressible fluids become incompressible. It is not hard to show that all reasonable types of corrections are scaled away,
and the incompressible Navier-Stokes equation universally governs the \hydrolimit~of essentially any fluid. The limit is an incredibly rich and interesting one because, even though the amplitudes are scaled to zero,  nonlinearities survive. It is this \hydrolimit~of a fluid that we will match to a near-horizon limit in gravity.
\section{Characterizing the dual geometries}

We seek a relation between the (p+2)-dimensional Einstein  and (p+1)-dimensional Navier-Stokes equations. Of course, the former has a much larger solution space than the latter so only a special type of Einstein geometry is relevant.  Roughly speaking, the relevant geometries are non-singular perturbations of a horizon. Let us now make this more precise.

\begin{figure}[h]
\begin{center}
\includegraphics[width=8cm]{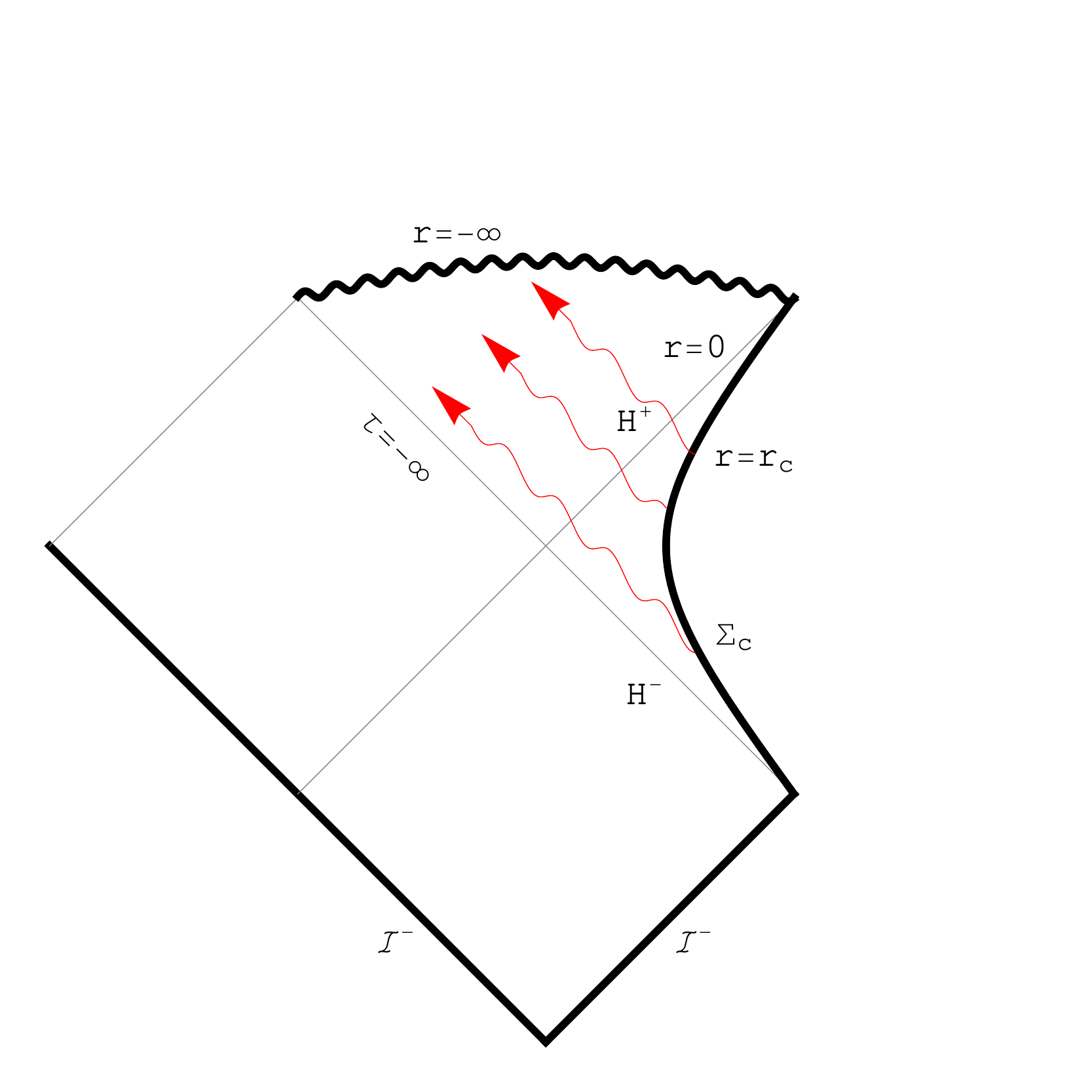}
\end{center}
\caption{This figure depicts the Einstein geometry holographically dual to a
fluid. The accelerated boundary hypersurface \sc\ at radius $r=r_c$ is
intrinsically flat but the extrinsic curvature is given by the fluid
stress tensor. This extrinsic curvature leads to gravity waves which
propagate radially inward. The leading-order condition that these
waves do not cross the past horizon \hm\ of \sc\ at $\tau=-\infty$ or
produce singularities on the future horizon \hp\ at $r=0$ is the non-linear incompressible Navier-Stokes equation for the fluid.}
\label{figure1}
\end{figure}

We consider geometries  of the type depicted in Figure \ref{figure1} with an  outer ``cutoff'' boundary denoted \sc.   The boundary hypersurface \sc\ is taken to be asymptotically null in both the far future and far past. In the Minkowskian coordinates
$ds_{p+2}^2=-dudv+dx_idx^i $, past null  infinity \im\ is the union of the null surfaces $v
\to-\infty$ together with $u \to-\infty$ and   \sc\ is the
timelike hypersurface $uv=-4 r_c$ with $v>0$. Past (future) event horizons \hm\ (\hp ) are defined by the boundaries of the causal future (past) of \sc.
The dual geometries  will be constructed in two a priori different expansions about Minkowski space: the near-horizon and the hydrodynamic $\e$-expansion.  Ultimately the two expansions will be shown to be equivalent.

Initial data can be  specified on the union  of \sc\ and \im .
We consider initial data which is asymptotically Minkowskian and flat (no incoming waves) on \im\  (or equivalently \hm).  On \sc\ we generally demand that the
intrinsic  metric $\gamma_{ab}$ be flat,
\beq  \gamma_{ab}=\eta_{ab},~~~a,b=0,...p \eeq  although we will later consider ``forcing''  the system by perturbing $\gamma_{ab}$.

We wish to consider the general solution of the Einstein equations consistent with this initial data and smooth on \hp.\footnote{Here we allow for incoming flux where \im\  meets \sc\ at $u=-\infty, ~v=0$.}
 In particular, so far we have not specified the extrinsic curvature $ K_{ab}$  on
\sc\ or equivalently (and more conveniently) the \brownyork\ stress tensor on \sc
\footnote{Our normalization here agrees with the conventional one for $G=1/16\pi.$}
\beq
T_{ab}  \equiv 2(\gamma_{ab} K -K_{ab} ).
\eeq
 If no initial data were prescribed on \im, any $ T_{ab}$ on \sc\ consistent with the constraint equations could be chosen.  This data could then in general be evolved radially inwards to produce a spacetime everywhere inside of \sc. In general, such a spacetime will have gravitational flux (if not singularities) going up to $v=\infty$ (${\cal I}^+$) as well as down to \im. Hence we have a ``shooting problem'' to find those special allowed choices of $ T_{ab}$ which produce a spacetime smooth on \hp\ with no flux coming up from \im.

We solved this problem in \cite{bkls} to leading order in a double expansion in long wavelengths and weak fields.\footnote{Our conventions here differ from \cite{bkls}.}
 Ingoing Rindler coordinates were used for which the leading order flat metric is
\beq
 \label{mcs} ds_{p+2}^2=-rd\tau^2+2d\tau dr+dx_idx^i.
\eeq
$\Sigma_c$ is the accelerated surface $r=r_c$, \hm\ is $\tau=-\infty$ and \hp\ is $r=0$.
These coordinates are convenient for analyzing smoothness on \hp.
It was found that the allowed choices of $T_{ab}$
are precisely those corresponding to the linearized fluid:
\beq
 r_c^{3/2} T^{\tau i}=v^i,~~~r_c^{3/2}T^{ij}=-\eta \p^{(i }v^{j)},
\eeq
where the (kinematic) viscosity here is given by the formula%
\beq\label{vis}
\eta= r_c,
\eeq
while $v_i$ obeys the linearized incompressible Navier-Stokes equation
\beq
\p_iv^i=0, ~~~\p_\tau v^i-\eta\p^2v^i=0.
\eeq
If we choose any value for the viscosity other than (\ref{vis}), the constraint equations on \sc\ are still obeyed, but gravitational waves are propagated down to \im\ and there is a singularity at $r=0$.

In this paper we go one step further and solve the problem in certain hydrodynamic and near-horizon limits  $without$  making a linearized approximation, enabling us to see a direct connection between the nonlinear structures of the Navier-Stokes and Einstein equations.

\section{Nonlinear solution in the $\epsilon$-expansion}

In this section we will improve on the analysis of \cite{bkls} by solving  the shooting problem in the long wavelength $\epsilon$-expansion without a simultaneous linearized expansion. The general  solution will be parameterized by a solution $v_i(x^i, \tau), \; P(x^i,\tau)$ of the full nonlinear Navier-Stokes equation with viscosity (\ref{vis}) together with the parameter $\epsilon$.

\subsection{The solution}
Consider the metric
\begin{align}
ds_{p+2}^2 =&-r d\tau^2 + 2 d\tau dr + dx_i dx^i \notag\\
&-2\left(1-\frac{r}{r_c}\right) v_i dx^i d\tau -2\frac{v_i}{r_c} dx^i dr \label{emet}\\
&+{\left(1-\frac{r}{r_c}\right)}\left[ (v^2+2P) d\tau^2  +\frac{v_i v_j}{r_c} dx^i dx^j \right] + \left( \frac{v^2}{r_c} +\frac{2P}{r_c}\right) d\tau dr   \notag\\
&-{\frac{(r^2 - r_c^2)}{r_c}}\p^2 v_i dx^i d\tau
+~\ldots\notag
\end{align}
where $v_i=v_i( x^i,\tau)$ and $P ( x^i, \tau)$
are independent of $r$.
Here and henceforth  $i,j=1,..p$ indices are raised and lowered with $\d_{ij}$ and we take
\beq\label{orders}
v_i\sim {\cal O}(\e),~~~~P \sim {\cal O}(\e^2),~~~~\p_i \sim {\cal O}(\e),~~~~\p_\tau \sim {\cal O}(\e^2)
\eeq
as in the hydrodynamic scaling of section 3.
It follows that the first line on the right hand side of (\ref{emet}) is ${\cal O}(\e^0)$ and each subsequent line is one higher order in $\e$. The linearization of this expression in $v_i$ agrees with the linearized solution studied in \cite{bkls}.

On the cutoff surface \sc, $r=r_c$ and the induced metric is flat:
\beq
\g_{ab}dx^a dx^b=-r_c d\tau^2+dx_idx^i,
\eeq
and hence satisfies the desired boundary condition.  Here and henceforth $x^a \sim (x^i, \tau)$.
The extrinsic curvature and unit normal on \sc\ are
\beq
 K_{ab}=\half \cl_N \g_{ab}=-\frac{1}{2}\left[T_{ab}-{1 \over p}\g_{ab}\g^{cd}T_{cd}\right],~~~~
N^\mu\p_\mu = {1\over \sqrt{r_c}}\p_\tau + \sqrt{r_c}\left(1 -\frac{P}{r_c}\right)\p_r +\frac{v^i}{\sqrt{r_c}}\p_i +\cO(\e^3).
\eeq
The \brownyork\ stress tensor is
\beq  \label{dss} T_{ab}dx^adx^b = {dx_i^2\over \sqrt{r_c}} +\frac{v^2}{\sqrt{r_c}} d\tau^2-2 \frac{v_i}{\sqrt{r_c}} dx^id\tau+\frac{(v_iv_j +P \delta_{ij})}{r_c^{3/2}} dx^idx^j  -2\frac{\p_iv_j}{\sqrt{r_c}}dx^idx^j
%
+{\cal O}(\e^3).
\eeq

We wish to solve the Einstein equations as a power series in $\e$.  We first consider the necessary but not sufficient condition that the constraints be satisfied on \sc. At order $\e^0$ the metric is flat and $T_{ab}$ is constant so they are trivially satisfied.   The only way to get an order $\e$ term is with one power of $v^i$ and no derivatives.
Such a linear term cannot appear because the constant $v^i$ terms in (\ref{emet}) can, through quadratic order, be obtained from a boost of flat space. The first nontrivial equation is encountered at order $\e^2$:
\beq
r_c^{3/2} \p_a T^{a\tau}=\p_iv^i=0 .
\eeq
This equation is satisfied if and only if  $v^i$ is the velocity field of an incompressible fluid. Taking this to be the case,  one finds at order $\e^3$:
\beq
r_c^{3/2} \p^a T_{ai}=\p_\tau v_i-\eta \p^2v_i+\p_iP+v^j\p_j v_i =0.
\eeq
This is satisfied if and only if $v^i$ solves the  Navier-Stokes equation with pressure $P$ and viscosity $\eta=r_c$.

Once the constraints are satisfied it is ordinarily possible to evolve the solution off the hypersurface, in this case in the radial direction, at least for a finite distance. Here we have the danger of
singularities at the horizon \hp\ near $r=0$, or equivalently waves coming up from \im. We know from \cite{bkls} that such singularities are absent in the linearized analysis provided the fluid viscosity takes the required value (\ref{vis}).  We have checked by direct computation that this absence of singularities extends to the nonlinear case as well.  That is all components obey
\beq
 \label{fcv} G_{ra},G_{ab},G_{rr}=\cO (\e^4)
\eeq
and are nonsingular for finite values of $r$.
Presumably the order $\e^4$ and higher terms in the metric can be chosen so that the Einstein equations are solved exactly.

It turns out that it is still possible to solve the Einstein equations analytically through order $\e^3$ with the ``wrong'' value of the viscosity (i.e. $\eta\neq r_c$ ) even in the nonlinear case. As expected these solutions develop a singularity at $r=0$ near \hp, and are presented in Appendix \ref{appendix}\ .
\subsection{Forcing the fluid}

Solutions of the linearized Navier-Stokes equation decay exponentially in the future.
There is some expectation - although no proof - that nonlinear solutions eventually decay as well. Therefore the extrinsic curvature on $\Sigma_c$ in our examples is expected to become constant.

On the other hand, already at the linear level, Navier-Stokes solutions grow exponentially in the far past and typically are singular at $\tau=-\infty$ . Therefore we expect that the dual geometry is also singular at $\tau=-\infty$, which is the past horizon \hm\ of $\Sigma_c$.
This singularity is not problematic for real fluids, as we are typically interested in cases where forcing terms correct the Navier-Stokes equation. For example we might consider a fluid which is initially at rest, stirred at time $\tau=\tau_*$ , and then left to evolve according to the unforced Navier-Stokes equation.

\begin{figure}[h]
\begin{center}
\includegraphics[width=8cm]{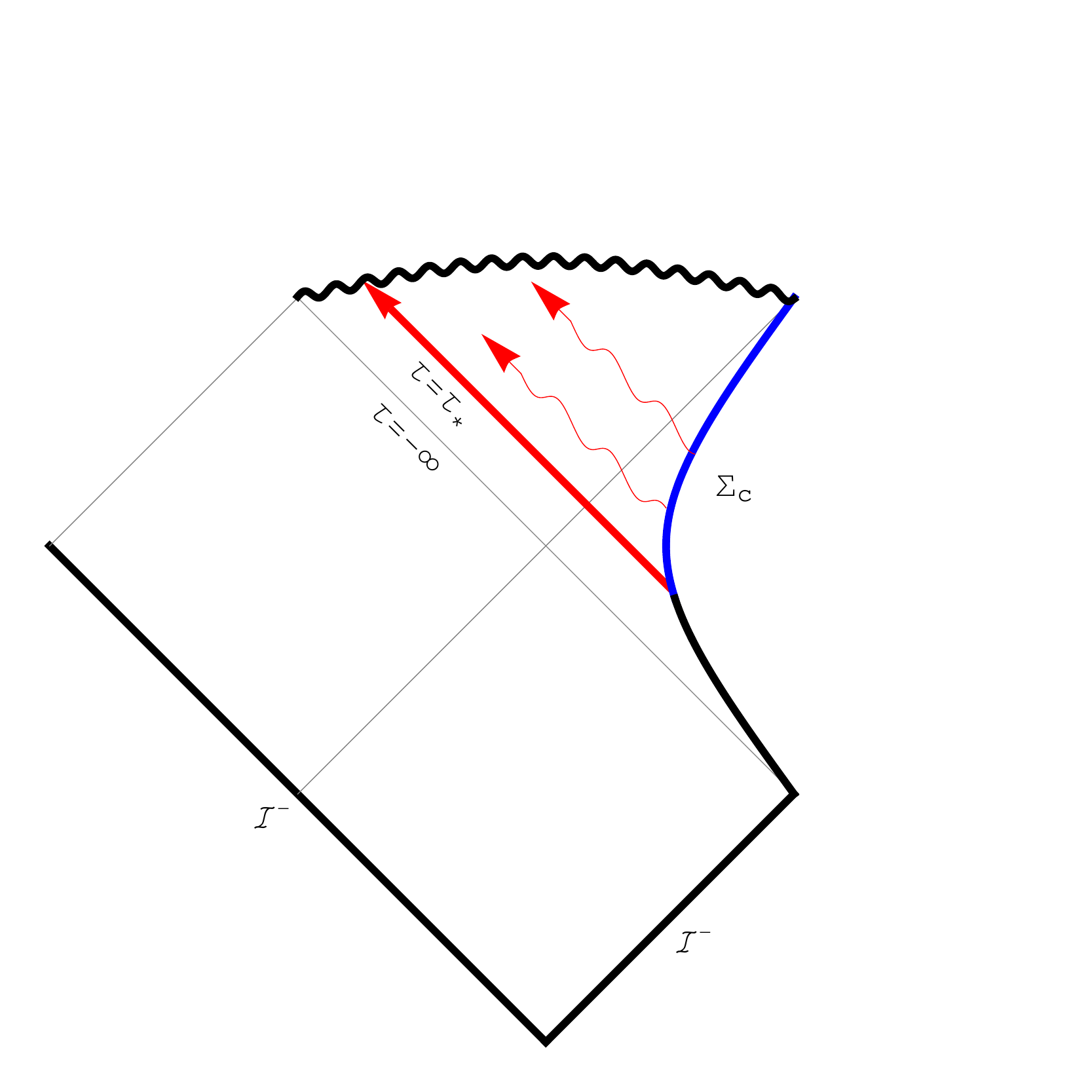}
\end{center}
\caption{On the \sc\ surface, prior to $\tau=\tau_*$, all initial data is trivial. At
$\tau=\tau_*$, a gravitational shock wave arrives. The shock forces the fluid
on \sc, and consequently the $v_i$ is nontrivial on \sc\ after $\tau_*$.
}
\label{figure2}
\end{figure}

In fact this  kind of situation is also very natural to consider on the gravity side. Consider flat Minkowski spacetime with a flat metric and constant extrinsic curvature on the boundary $\Sigma_c$ for $\tau < \tau_*$. We then stir it at $\tau=\tau_*$ by momentarily perturbing the boundary condition that  the induced metric on $\Sigma_c$ be flat. This will send out a gravitational shock wave along $\tau=\tau_*$ and excite the geometry for $\tau >\tau_*$.  The result should be an appropriate gluing of (\ref{emet}) along a null hypersurface to flat Minkowski space. This is depicted in Figure \ref{figure2}.

At the linear level, it is possible to explicitly construct the glued geometry describing this situation through order $\e^3$. The metric is
\begin{align}
ds^2 =&-r d\tau^2 + 2 d\tau dr + dx_idx^i \notag\\
&-\left[ 2 (1 - r/r_c)  v_i dx^i d\tau +  (1-r/r_c) \left(\p_jv_i + \p_i v_j \right) dx^i dx^j - 2 \left(r - \frac{r^2}{ 2 r_c} - r_c/2 \right)\p^2 v_i dx^i d\tau \right]\notag\\
&-\delta(\tau-\tau_*)\left[\left( 4 (1 - r/r_c)  F_i  + \frac{2}{r_c } \alpha_{i} \right) dx^i d\tau - \frac{2  }{r_c}  \beta_{ij} dx^i dx^j \right] + \ldots\label{rrt}
\end{align}
where $F_i$ is an arbitrary function of $x^i$ obeying $\p_iF^i=0$.  $\beta_{ij}$ and $\alpha_{i}$ (which is divergence free) are both functions of $x^i$ and related to $F_i$ by
\beq
\p^j\p_j \alpha_{i}=F_i,\qquad \beta_{ij} = \p_i \alpha_j + \p_j \alpha_i.
\eeq
Since the metric on $\Sigma_c$ is no longer flat, the constraint equations become linearized Navier-Stokes with a forcing term similar to that described in \cite{bkls}. For this configuration we have
\beq \label{oi}
\p_\tau v^i-\eta\p^2v^i=F^i(x)\delta(\tau-\tau_*).
\eeq
Clearly, since $v_i(x, \tau)$ is taken to vanish for $\tau<\tau_*$, the forcing term will cause it to jump to $F_i(x^i)$ at $\tau=\tau_*$, after which it will evolve according to Navier-Stokes.
Given (\ref{oi}) this geometry solves the linearized Einstein equations everywhere, and is characterized by an arbitrary divergence-free vector field $F_i(x)$. Before $\tau=\tau*$ it is flat, while afterward it is, up to a coordinate transformation,  the linearization of (\ref{emet}).

At the nonlinear level, the equations are cumbersome and we have been unable to explicitly construct the analog of (\ref{rrt}) away from $\Sigma_c$.  However it seems plausible that qualitatively similar solutions persist at the nonlinear level.
\subsection{Singularities at $r=\pm\infty$}

The square of the Riemann tensor for the solution  (\ref{emet}) is given by
\beq \mathcal{R}^2 =- \frac{3}{2 r_c^2} \left(\p_i v_j - \p_j v_i \right)^2 - 2 \frac{ r}{r_c^2} \left[ \p^2 v_i \p^2 v^i + 3 \p^i v^j \left(\p_j \p^2 v_i - \p_i \p^2 v_j \right) \right] +~\ldots.
\eeq
This expression diverges at  $r=\pm \infty$. Of course perturbation theory cannot be trusted
when $|r|$ is of order ${1 \over \e}$, so the computation is unreliable  in this regime.
Whether or not there are actual divergences in these regions will depend on the details of the solution. In general, at $r=-\infty$, black hole type singularities may plausibly arise.

The divergence at $r=+\infty$ is outside the cutoff surface, so a priori need not concern us.
 Still we may ask
what happens if we try to extend the solution to this region. In general relativity with no cosmological constant it is hard to find solutions which are asymptotically flat in codimension one: i.e. there are no codimension one black holes.  This suggests that many configurations will be singular if extended  to $r=+\infty$.  On the other hand, if we add a negative cosmological constant,  there are codimension one asymptotically AdS black holes.  At large $r$ the cosmological term tends to dominate, and we expect in this case many solutions to have nonsingular extensions to this region. However, as we will see below, the hydrodynamic regime is small $r$ so the large $r$ behavior is of limited interest for the present purposes.

\subsection{Uniqueness}
  Equation (\ref{emet}) gives the first three orders in the $\e$-expansion of metrics satisfying the Einstein equations with the prescribed boundary data. These solutions are constructed from nonlinear solutions of the incompressible Navier-Stokes equations.  The latter are in turn, assuming existence and uniqueness for Navier-Stokes, specified by a divergence-free vector field $v^i(x,\tau_*)$ at a moment of time $\tau_*$.

    One may ask whether or not (\ref{emet}) is the unique solution with the prescribed  boundary data (up to coordinate transformations and field redefinitions)  associated to a given
$v^i(x,\tau_*)$. This can be addressed in the context of a combined weak-field expansion and $\e$-expansion. The problem was solved to leading nontrivial order in the weak-field expansion in \cite{bkls}. The unique solution is the first two lines of (\ref{emet}), but with a $v^i$ obeying the linearized Navier-Stokes equation. Generally one does not expect
the dimension of the solution space in weak-field perturbation theory to change unless there is a linearization instability and associated obstruction. In the present case, the only potential obstruction is the Navier-Stokes equation which we are assuming can be solved.
Hence one expects the solution (\ref{emet}) to be unique at each order in the $\e$-expansion, up to the usual ambiguity of adding solutions of the leading order equations at subleading orders.
\section{Alternate presentation}
In this section we give an alternate presentation of the  metric (\ref{emet}) in which all the factors of $\e$ appear explicitly, without being hidden in the functional dependence on the coordinates. This is accomplished by first transforming to hatted coordinates
\beq
{x^i }  ={r_c \x^i \over \e}, ~~~~ \tau= { r_c \t \over \e^2}, ~~~~r= r_c\r
\eeq
so that $\p_{\t }=\cO(\e^0)$ and we denote $\hat \p_i={\p \over \p \x^i}=\cO(\e^0)$. In the new coordinates
\begin{align}
ds_{p+2}^2 =&-\frac{\r  r_c^3}{\e^4} d\t^2 + \frac{2r_c^2}{ \e^2} d\t d\r + \frac{r_c^2 }{ \e^2}d\x_i d\x^i \notag\\
&-2 r_c^2\frac{1-\r}{\e^2}\v_i d\x^i d\t -2 r_c{\v_i}d\x^i d\r \label{emts}\\
&+{(1-\r)}\left[r_c^2 \frac{ \v^2+2\P}{\e^2} d\t^2 + r_c{\v_i \v_j }d\x^i d\x^j \right]  + r_c(\v^2+2\P)d\t d\r \notag\\
&-( \r^2-1)r_c \hat \p^2 \v_i d\x^i d\t +~\ldots,
\end{align}
where $\P(\x,\t)={1 \over\e^2}P(x(\x),\tau(\t))$, $\v_i(\x,\t)={1\over \e}v_i(x(\x),\tau(\t))$,
$\v^2\equiv \v_i\d^{ij}\v_j$ and $i,j$ indices are raised and lowered with $\d_{ij}$.  The usual Navier-Stokes equation for $v,~P$ with $\eta=r_c$ implies
\beq
\p_{\t} \v_j-\hat \p^2\v_j+\v^k\hat \p_k\v_j+ \hat \p_j\P =0.
\eeq
This is the Navier-Stokes equation with $\eta=1$ and no factors of $\e$ or  $r_c$.

Finally let us consider the rescaled metric ${d\hat s}_{p+2}^2={\e^2 \over r_c^2} ds_{p+2}^2$.
The Einstein tensor is invariant under such constant metric rescalings. Rearranging terms and defining
\be \lambda\equiv \frac{ \e^2}{r_c} \ee
one finds
\begin{align}
d\hat s_{p+2}^2 =&-{\r \over \lambda} d\t^2 \notag\\
&+\bigl[2d\t d\r + d\x_i d\x^i-2 (1-\r)\v_i d\x^i d\t +{(1-\r)} ( \v^2+2\P) d\t^2 \bigr]  \label{mts}\\
&+\lambda\bigl[(1-\r){\v_i \v_j }d\x^i d\x^j   -2 {\v_i}d\x^i d\r + (\v^2+2\P)d\t d\r +(1-\r^2)\hat \p^2 \v_i d\x^i d\t
\bigr]
~+\ldots.\notag
\end{align}
The \brownyork\ stress tensor is
\beq
\left\{\hat T^{\t}_{\t} = - \sqrt{\lambda} \v^2 ,\qquad \hat T^{\t}_i = - \sqrt{\lambda} \v_i,\qquad \hat T^i_j = \frac{1}{\sqrt{\lambda}}\delta^i_j  + \sqrt{\lambda} \left[ \v^i \v_j + \P \delta^i_j - 2\hat \p^i \v_j  \right]\right\}\qquad + \cO(\lambda^{3/2} ).
\eeq
The important point here is that the geometry depends only on the ratio $\lambda={\e^2\over r_c}$ and not $\e$ or $r_c$ separately.

Given that the rescaled geometry depends only on $\lambda$ and the $\e$-dependence (\ref{fcv}) of the unrescaled geometry (\ref{emet})
we conclude that in the hatted coordinates
\begin{align}
G_{\t \t}&= {r_c^2 \over \e^4} G_{\tau\tau}\sim \cO(\lambda^{0}),
\quad G_{\hat{i}\hat{j}}= {r_c^2\over \e^2} G_{ij}\sim \cO(\lambda),\notag\\
\quad G_{\r\t}&= {r_c^2 \over \e^2}G_{rr}\sim \cO(\lambda),
\quad G_{\r\r}= r_c^2G_{rr}\sim \cO(\lambda^2) \notag\\
G_{\t \hat{i}}&= {r_c^2 \over \e^3} G_{\tau i}\sim \cO(\lambda^{1/2}),
\quad G_{\r \hat{i}}= {r_c^2 \over \e} G_{r i}\sim \cO(\lambda^{3/2}).
\label{nhg}
\end{align}
Given the explicit factor of $\lambda^{-1}$ in $g_{\t\t}$, it is not immediately obvious in this presentation that in a direct computation the Einstein tensor will even have a good Taylor expansion in $\lambda$. What happens is that, because $g_{\r\r}=0$, there are only a limited number of powers of $g_{\t\t}$ that can appear in the Einstein tensor, and one may thereby directly recover (\ref{nhg}).  In fact, direct computation reveals  we do slightly better; the last line may be replaced by
\beq
G_{\t \hat{i}}= {r_c^2 \over \e^3} G_{\tau i}\sim \cO(\lambda^{1}),
\quad G_{\r \hat{i}}= {r_c^2 \over \e} G_{r i}\sim \cO(\lambda^{2}).
\eeq

Notice that $G_{\t\t}$ in (\ref{nhg}) is of order $\lambda^0$ rather than $\lambda^1$.  We can improve this by computing a few higher order pieces of the metric.  Specifically, we add to  (\ref{mts})
\beq\label{ellipsisdef}
-2\lambda (1-\r)\q_{i} d\x^i d\t  +2 \lambda^2 g^{(2)}_{\hat{r}i} d\r d\hat{x}^{i}+\lambda^2 g^{(2)}_{ij} d\hat{x}^{i}d\hat{x}^{j}+~\ldots
\eeq
Demanding that the $r$-independent pieces of $G_{\t\t}=0$ be solved through order $\lambda^0$ then fixes $\q_i(\t,\hat{x})$:
\beq
 \label{op} \hat \p_i\q^i= \hat\p^2 \v^2 -  \frac{1}{2}\v^i  \hat \p_i  \v^2  - \frac{3}{2} \p_{\t} \v^2-\frac{1}{2} \left( \hat \p_i \v_j + \hat \p_j \v_i \right)^2 .
 \eeq
Apparently $\q_i$ is  a kind of heat current.
Demanding that the entire $G_{\t\t}=0$ through order $\lambda^0$ gives us a differential equation for the combination $\hat Q(\r,\t,\hat{x})\equiv - 2 \hat \p^{i}  g_{{\r}i}^{(2)} + \p_{\r}g_{~i}^{i~{(2)}}$:
\beq
\label{nyqe}
\hat Q + 2 \r \p_{\r} \hat Q =   2\hat \p_i \q^i - 2 \v_i \q^i  + 3 \r \hat \p^2 \v^2 -  \frac{\r}{2} \left(\hat \p_i \v_j +\hat \p_j \v_i \right)^2 + 2 \hat \p_i \v_j  \hat \p^j \v^i  -  \v^j \hat \p_j \v^2 + \left(\v^2 \right)^2 - 2 \v^i \hat \p_i \hat P + 2 \hat P \v^2 .
\eeq
Choosing $\q_i,~\hat Q$ accordingly, we find that as desired all components of the Einstein equations vanish for $\lambda \to 0$:
\beq
\label{fcdv} G_{\r\hat a},G_{\hat a \hat b},G_{\r\r}=\cO(\lambda).
\eeq
\section{Petrov type}
Interestingly, this geometry is of an algebraically special type. We consider the case of $p=2$ to connect to the well-studied Petrov classification of spacetimes \cite{exac}). A geometry is Petrov type II if there exists a real null vector $k^\mu$ such that the Weyl tensor satisfies
\beq\label{weyco}
W_{\mu\nu\rho[\sigma} k_{\lambda]}k^\nu k^\rho=0.
\eeq
This happens if the invariant $I^3 - 27 J^2$ vanishes where $I,J$ are both specific combinations of Weyl tensor components which can be found in  \cite{exac}. For the metric (\ref{emet}), the lowest nonzero entries for $I,J$ are at $\cO(\epsilon^4)$ and $\cO(\epsilon^6)$ respectively.  Hence, the first contribution to the invariant would be at $\cO(\epsilon^{12})$; however the invariant vanishes through $\cO(\e^{13})$. At higher order in $\e$, it gets modified by corrections to (\ref{emet}); we expect that including higher order terms in (\ref{emet}) enables (\ref{weyco}) to be satisfied exactly.

\section{Nonlinear solution in the near-horizon expansion}

In section 4, the nonlinear Einstein equations with certain boundary conditions were solved in the non-relativistic, long-wavelength hydrodynamic $\e$-expansion.  This generalized the analysis given in \cite{bkls} of the $\e$-expansion for linearized modes.   \cite{bkls} also considered, for linearized modes, a second, near-horizon expansion. Although physically inequivalent, the two expansions were found to be equivalent mathematically and reduce to the linearized dynamics of an incompressible fluid. In this section, we consider the nonlinear version of the near-horizon expansion and find that it is again mathematically equivalent to the nonlinear $\e$-expansion.

In the $\e$-expansion one solves the shooting problem for long-wavelength perturbations of \sc\ with a fixed leading-order extrinsic curvature.
The proper acceleration of a worldline at fixed $x^i$ in \sc\ is to leading order just proportional to $K_{\tau\tau}$, so we may also view this as fixing the  acceleration  away from the origin.  In the
near-horizon expansion, instead of expanding in the wavelength one expands in the inverse acceleration. We begin with the flat metric on the Rindler wedge
\beq
ds_{p+2}^2 =-r d\tau^2 + 2 d\tau dr + dx_i dx^i .
\eeq
To avoid confusion with the notation of the previous section we put the boundary on the accelerating surface
\beq
 r=\tilde r_c,
\eeq
so that $r\le \tilde r_c$.
The near-horizon, large acceleration,  limit is
$\tilde r_c\to 0$. In order to exhibit the $\tilde r_c$-dependence explicitly in the metric we transform to $r=\tilde r_c\r$, $\tau= {\t \over \tilde r_c}$ so that $r\le 1 $ and
\beq
ds_{p+2}^2 =-{\r \over \tilde r_c} d\t^2 + 2 d\t d\r + dx_i dx^i .
\eeq
In these coordinates the near-horizon limit rescales to infinity the coefficient of $d\t^2$ at any finite $\r$.

We now wish to consider perturbations of this metric solving the Einstein equations order by order in the near-horizon expansion parameter $\tilde r_c$ that are  consistent with a flat induced metric at
$\r=1$. At the level of linear perturbations, the most general solution was found in \cite{bkls} (characterized in terms of the data at $r=\tilde r_c$).
This solution is (for all $r$)
\beq\label{xzmts}
d\hat s_{p+2}^2 =-{\r \over \tilde r_c} d\t^2 +2d\t d\r + dx_i dx^i-2 (1-\r)v_i dx^i d\t   + \tilde r_c \left[(1-\r^2) \p^2  v_i dx^i d\t-2   { v_i}dx^i d\r \right]+ {\cal O}(\tilde r_c^2)
\eeq
where $\p_iv^i=0$ and $\p_{\t } v^i- \p^2 v^i=0$. That is, $v^i$ is an incompressible fluid flow obeying the linearized Navier-Stokes equation with unit kinematic viscosity.

The nonlinear generalization of (\ref{xzmts}) which solves the nonlinear Einstein equations to
$\cO(\tilde r_c)$ is
\begin{align}
d\hat s_{p+2}^2&=-{\r \over \tilde r_c} d\t^2\notag\\
&+\bigl[2d\t d\r + dx_i dx^i-2 (1-\r) v_i dx^i d\t +{(1-\r)} (  v^2+2\P) d\t^2 \bigr]  \notag\\
&+\tilde r_c\bigl[(1-\r){ v_i  v_j }dx^i dx^j   -2 { v_i}dx^i d\r + ( v^2+2\P)d\t d\r +(1-\r^2) \p^2  v_i dx^i d\t-2 (1-\r) \q_i(\t,\r,x) dx^i d\t\bigr] \notag\\
&+ \tilde r_c^2 \left[ 2 g_{{\r}i}^{(2)}(\t,\r,x) dx^i d\r + g_{ij}^{(2)}(\t,\r,x) dx^i dx^j \right] + {\cal O}(\tilde r_c^2)\label{mtxs}
\end{align}
provided $ \p_iv^i=0$, $ \p_{\t} v_j- \p^2 v_j+ v^k \p_k v_j+  \p_j \P =0$.   $\q_i,~ g_{{\r}i}^{(2)},~g_{ij}^{(2)}$ are solutions of  first order differential equations of the type (\ref{op}) and (\ref{nyqe}). Further ${\cal O}(\tilde r_c^2)$ pieces do not affect the equations of motion to this order.

We can now see explicitly that making the notation change $v\to \hat v, ~x^i\to\x^i$ and $\tilde r_c \to \lambda$
in (\ref{mtxs}) gives us the rescaled solution (\ref{mts}) in section 6. Hence the near-horizon and hydrodynamic expansions are mathematically equivalent.

Since we are identifying $\tilde r_c=\lambda={\e^2\over r_c}$, $r_c\to \infty$ in the metric (\ref{emet}) is actually equivalent to the  near-horizon limit $\tilde r_c\to 0$ in (\ref{mtxs}).  This may at first seem odd, but the near-horizon-hydrodynamic equivalence involves a constant rescaling of (\ref{emet}) by a factor of ${1 \over r_c^2}$, the proper distance to the cutoff surface in the rescaled metric (\ref{mtxs}) indeed behaves as  $1 \over \sqrt{r_c}$.
\section{Acknowledgements}
We are grateful to D. Christodoulou, S. Cremonini, T. Damour, B. Freivogel, S. Gubser, S. Hartnoll, S. Kachru, R. Loganayagam, A. Maloney, S. Minwalla, D. Nelson, P. Petrov, S. Sachdev, O. Saremi, J. Smoller, K. Thorne, E. Verlinde and S. T. Yau for illuminating conversations. This work was supported by DOE grant DE-FG0291ER40654 and the Fundamental Laws Initiative at Harvard.

\appendix
\section{Appendix}\label{appendix}

In the $\epsilon$-expansion,
\begin{align}
ds_{p+2}^2 =&-r d\tau^2 + 2 d\tau dr + dx_i dx^i -2\left(1-\frac{r}{r_c}\right) v_i dx^i d\tau -2\frac{v_i}{r_c} dx^i dr \notag\\
&+{\left(1-\frac{r}{r_c}\right)}\left[ (v^2+2P) d\tau^2  +\frac{v_i v_j}{r_c} dx^i dx^j \right] + \left( \frac{v^2}{r_c} +\frac{2P}{r_c}\right) d\tau dr + c_1 \log\left(\frac{r}{r_c}\right) \left( \p_i v_j + \p_j v_i \right)dx^i dx^j  \notag\\
&-{\frac{(r^2 - r_c^2)}{r_c}}\p^2 v_i dx^i d\tau -2\left(1-\frac{r}{r_c}\right) q_i dx^i d\tau - 2 c_1 \left( r \log r - r_c \log r_c \right) \p^2 v_i dx^i d\tau ~\label{wrdmet}\\
&- 2 c_1  \log\left(\frac{r}{r_c}\right) v^j \left(\p_i v_j + \p_j v_i \right) dx^i d\tau + 2 c_1 \left(1 - \frac{r}{r_c} \right) v^j \p_j v_i dx^i d\tau+{\cal O}(\e^4)\notag
\end{align}
solves the Einstein equations through ${\cal O}(\e^3)$ if $v_i$ obeys the incompressible Navier-Stokes equation with the ``wrong'' viscosity $\eta=r_c \left(1 + c_1 \right)$ where $c_1$ is a nonzero constant. For this geometry, the square of the Riemann tensor is
\beq\label{wrdriesq}
\mathcal{R}^2=- \frac{3}{2 r_c^2} \left(\p_i v_j - \p_j v_i \right)^2 + \frac{c_1 \left(c_1 + 2\right)}{r^2} \left[ 2 \p_i v_j \p^j v^i + \frac{1}{2} \left(\p_i v_j - \p_j v_i \right)^2 \right] + O(\epsilon^6)
\eeq
which clearly diverges at $r=0$ unless $c_1$ vanishes or $c_1=-2$. The last possibility is the time reverse of the first and exponentially growing in the future.

\bibliography{ref}{}
\bibliographystyle{utphys}
\end{document}